# Hybrid-graph neural network method for muon fast reconstruction in neutrino telescopes

Cen Mo and Liang Li[*]

*State Key Laboratory of Dark Matter Physics, Key Laboratory for Particle Astrophysics and Cosmology (MOE), Shanghai Key Laboratory for Particle Physics and Cosmology (SKLPPC), School of Physics and Astronomy, Shanghai Jiao Tong University, Shanghai 200240, China*

Fast and accurate muon reconstruction is crucial for neutrino telescopes to improve experimental sensitivity and enable online triggering. This paper introduces a hybrid-graph neural network (GNN) method tailored for efficient muon track reconstruction, leveraging the robustness of GNNs, alongside traditional physics-based approaches. The "light GNN model" achieves a run-time of 0.19–0.29 ms per event on GPUs, offering a 3 orders of magnitude speedup compared to traditional likelihood-based methods, while maintaining a high reconstruction accuracy. For high-energy muons (10–100 TeV), the median angular error is approximately 0.1°, with errors in reconstructed Cherenkov photon emission positions being below 3–5 m, depending on the GNN model used. Furthermore, the semi-GNN method offers a mechanism to assess the quality of event reconstruction, enabling the identification and exclusion of poorly reconstructed events. These results establish the GNN-based approach as a promising solution for next-generation neutrino telescope data reconstruction.

## I. INTRODUCTION

Neutrinos serve as novel messengers for astrophysical research. Owing to their neutral charge and weak interaction strength, neutrinos can traverse vast cosmic distances without being deflected or absorbed by intervening matter. By detecting the Cherenkov radiation emitted by secondary particles produced in neutrino interactions, these telescopes enable the exploration of astrophysical neutrino's energy and direction, offering valuable tools to unravel the mysteries surrounding the origins of cosmic rays.

IceCube unveiled the first evidence of high-energy extraterrestrial neutrinos in 2013 [1]. Building on IceCube's groundbreaking discoveries, several advanced neutrino telescopes are currently in development, including the Tropical Deep-Sea Neutrino Telescope (TRIDENT) [2], KM3NeT [3], and Baikal-GVD [4]. Positioned in the South China Sea, TRIDENT is set to be a next-generation neutrino observatory, distinguished by its innovative design and substantial sensitive volume, which are expected to significantly enhance the detection capabilities for astrophysical neutrino sources.

Muon neutrinos are particularly important due to their fine angular resolution. High-energy muons are produced through charged-current interactions. As these muons pass through media such as seawater, they leave long tracks, which can be used to infer the original neutrino's direction. The average muon energy loss can be described by the equation $\langle -dE/dx \rangle = a + bE$ [5], where for seawater, $a = 2.6\ \mathrm{MeV\,g^{-1}\,cm^2}$ and $b = 3.6 \times 10^{-6}\ \mathrm{g^{-1}\,cm^2}$ [6]. This indicates that a 1 TeV muon can typically traverse 2.4 km, providing a long track, enhancing the angular resolution for muon direction reconstruction. Muon neutrinos were instrumental in the detection of neutrino emissions from TXS 0506 [7] and NGC 1068 [8] by IceCube. Furthermore, the recent identification of ultra-high-energy cosmic neutrinos by KM3NeT [9] was also achieved through a muon event.

Atmospheric muons are generated from interactions between cosmic rays and the atmosphere, with a high cosmic ray intensity leading to a substantial atmospheric muon flux. This makes atmospheric muons a crucial background for neutrino telescopes. For TRIDENT, a large-scale neutrino telescope to be located at a depth 3 km below sea level, the atmospheric muon flux with energies greater than 100 GeV is estimated at the order of 2500 Hz at the distance 2.7 km below sea level. In contrast, the rate of detectable atmospheric muon neutrinos is approximately an order of 0.01 Hz, with only a few tens of signal astrophysical neutrinos expected to be detected annually.

[*]Contact author: liangliphy@sjtu.edu.cn

To effectively distinguish neutrino events of interest from the overwhelming background of atmospheric muons, as well as to enhance sensitivity of neutrino telescopes, rapid and precise algorithms for the reconstruction of muon track direction is essential.

While likelihood-based methods can achieve high-resolution results, the minimization process is often time consuming, and the likelihood function requires manual tuning of model parameters from calibrated data. To address these challenges, the IceCube experiment explored the use of neural network techniques such as sparse submanifold convolutional neural networks [10] and recursive neural networks [11] for fast, trigger-level reconstruction. These neural network methods require large iteration only during the training phase. Once a model is trained, the reconstruction time depends solely on the number of model parameters and input data, making neural networks a promising approach for reconstructing muon events in neutrino telescopes.

This paper presents a novel method for the rapid and precise reconstruction of muon tracks in neutrino telescopes, combining graph neural networks (GNNs) with the least-squares method. Unlike other techniques that use neural networks to directly reconstruct muon direction, our approach utilizes GNNs to estimate intermediate variables—specifically, the injection positions of photons. These positions are then used in a least-squares method for a straightforward line-fitting process to derive the final results. The GNN model is expected to leverage the inherent sparsity and connectivity of the data, enabling fast and accurate inferences, while integration of the least-squares method enhances the stability and precision of the outcomes. Furthermore, this methodology includes a mechanism to identify and exclude failed reconstructions, adding an extra layer of reliability to the reconstruction process.

The remaining sections of this paper are organized as follows. Section II presents the simulated muon samples within the TRIDENT detector used in this study. In Sec. III, we introduce the reconstruction strategy. Section IV presents the performance of the designed method and discusses approaches for sanity checks on model output. Section V illustrates the impact of noise to the reconstruction accuracy. Summary and outlook are given in Sec. VI.

## II. MUON SAMPLE PREPARATION

Figure 1 shows the TRIDENT array used in this study. The array consists of about 1000 strings, each equipped with 20 hybrid digital optical modules (hDOMs) spaced 30 m apart vertically. Each hDOM comprises 31 photomultiplier tubes (PMTs) and 24 silicon photomultipliers (SiPMs), which are the primary units employed to detect signal photons. The arrangement of the strings follows a Penrose tiling distribution with interstring distances of 70 and 110 m [2]. The entire array occupies a cylindrical area with a radius of roughly 2000 m and a height of 700 m.

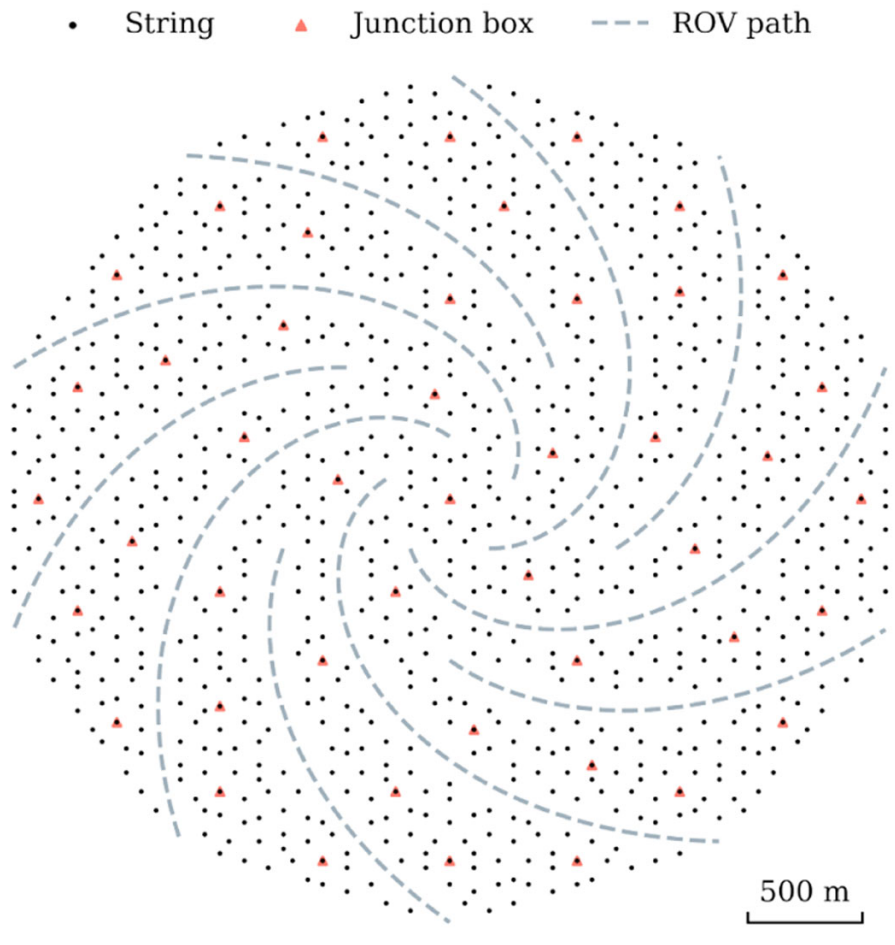

FIG. 1. Top-down view of the TRIDENT array. The geometrical layout pattern follows a Penrose tiling distribution. Each black dot represents a string of length ~0.7 km and the dashed lines highlight the paths designed for underwater maintenance by remotely operated underwater vehicles (ROVs) [2].

To simulate the response of the TRIDENT array to muons, we inject muons from the surface of the array. Three sets of simulations are conducted for muon energy [0.1, 1], [1, 10], and [10, 100] TeV separately. Within each set, muon energies are sampled from a spectrum $f(E) = E^{-1}$ to ensure a uniform distribution in $\log(E)$. This method is chosen to equalize the statistical weight of muons across a wide range of energies. Additionally, to achieve isotropic coverage, their momentum directions are uniformly distributed across the $4\pi$ solid angle. To determine the injected position of a muon, a random point within the TRIDENT array is chosen. From here, a half line opposite to the muon momentum direction is drawn to find its intersection with the surface of the TRIDENT array. The intersection is marked as the injection point. This procedure ensures the intersection between muon track and the TRIDENT array.

The Geant4 software framework [12–14] is used to simulate muon propagation, their interaction in water, subsequent particle production, and Cherenkov photon emission. Following this, the Optix ray tracing framework [15], which utilizes GPU acceleration, is employed for the propagation of the produced Cherenkov photons [16]. Photons that land on with PMTs or SiPMs are recorded by the corresponding digital optical module and PMT/SiPM IDs along with their arrival times.

To model detector response, we consider the wavelength-dependent quantum and collection efficiency of both PMTs and SiPMs [2]. Additionally, the transient time spread (TTS) of the PMTs and the jitter time of SiPMs are simulated by assuming a Gaussian-distributed time uncertainty with a standard deviation of $\sigma = \frac{\mathrm{TTS/jitter}}{\sqrt{2\ln 2}}$. For this study, we set TTS $= 3$ ns and jitter $= 0.1$ ns for PMTs and SiPMs, respectively.

In this study, $9 \times 10^5$ events are simulated. Among them, 150,000 events (50,000 events in each energy region) are isolated as a test set, to provide the results given in Sec. IV. The remaining events are allocated to the training and validation sets in a 10:1 ratio. As a preliminary trigger criteria, we only retain events where more than 15 photons are detected by at least two hDOMs.

## III. MUON TRACK RECONSTRUCTION METHOD

### A. Likelihood reconstruction

The arrival times and positions of the first photon of each hDOM are used to reconstruct the original muon direction. The method in processing this information strongly affects the reconstruction speed and accuracy.

The maximum likelihood method is commonly used in neutrino telescopes to reconstruct muon events. It utilizes a probability density function (PDF), $p(t_{\mathrm{res}}, d|\mathbf{v}, \mathbf{n})$, to calculate the likelihood for each muon event [17]. Here, $t_{\mathrm{res}}$ is defined as the time difference between the expected and measured arrival time of photon and $d$ represents the distance between the triggered hDOM and the muon track. $\mathbf{v}$ and $\mathbf{n}$ are the initial position vector and direction vector of the muon track, respectively. This method can typically achieve accurate reconstruction of direction and position after maximizing likelihoods. However, the iterations needed in maximizing the likelihood expression leads to a large run-time cost, especially when there are many photons received.

### B. Convolutional neural network reconstruction

Muon events can be viewed as images and be reconstructed using convolutional neural networks (CNNs). By dividing the detector array into cubic grids in three-dimensional (3D) space and aggregating the detector response within each grid, muon events can be transformed into 3D images, where the time and charge act as the color channels of the images. Employing CNNs to reconstruct these event images appears promising, given the significant success of CNNs in image-based tasks. This approach has been studied for various applications in neutrino telescopes, such as cascade reconstruction [18,19]. However, certain aspects of neutrino telescopes limit the effectiveness of CNNs [10]. Assuming 1 pixel/hDOM, an image representing the response of the TRIDENT detector contains 72,000 pixels. However, for muons of energy less than 100 TeV, only $\mathcal{O}(10)$ hDOMs typically detect photons, corresponding to an occupancy of less than 0.1%. This sparse representation is highly computationally inefficient and poses significant challenges in training CNNs. Moreover, the combination of sparsity and high resolution reduces the effectiveness of convolution operations. These operations are primarily designed to aggregate localized information, but in sparse images, the information is typically concentrated in a few isolated regions, rendering traditional convolutions suboptimal.

### C. GNN reconstruction

In this study, we represent muon events as edgeless graphs and employ GNNs for event reconstruction. Each muon event can be represented as a collection of hDOMs that detect photons. This leads to a graph-based representation, $\mathcal{G} = \{\mathcal{V}\}$, where $\{\mathcal{V}\}$ consists of hDOMs, serving as the nodes (vertices) of the graph $\mathcal{G}$. This representation eliminates unnecessary computation and storage for untriggered hDOMs, significantly enhancing computational efficiency. Moreover, by treating hDOMs as the fundamental unit in this representation, this approach is independent of the array's geometry, making it applicable to various neutrino telescopes. GNNs extract hidden knowledge of graphs by transferring effective information among nodes, where each transfer utilizes features of several nodes. This ensures that GNNs can achieve high reconstruction efficiency with respect to computing power.

For input to the GNN, each node is characterized by two attributes: a position vector and a feature vector. The position vector (**pos**) is a three-dimensional spatial vector indicating the relative position of an hDOM with respect to the first triggered hDOM. The feature vector $\mathbf{x}$ is a six-dimensional vector, $[||\mathbf{pos}||, \mathrm{nhits}, t, \mathbf{pos}]$, where "nhits" represents the number of received photons (limited to 100), and $t$ denotes the arrival time of the first detected photon at the corresponding hDOM.

Consequently, each muon event is represented by a graph whose nodes encompass both position and feature vectors: $\mathcal{G} = \{\mathbf{pos}, \mathbf{x} \equiv [\mathrm{nhits}, ||\mathbf{pos}||, t, \mathbf{pos}]\}$. The edges of graphs are dynamically established by the GNN and are not explicitly included in the initial definition of the graph.

### D. GNN architecture

Reference [20] introduces edge convolution (EdgeConv), an effective method for incorporating local neighborhood information within a point cloud. In this context, a point cloud is defined by a graph $g = \{\mathbf{pos}, \mathbf{x} \equiv \mathbf{pos}\}$. EdgeConv first constructs edges $\mathcal{E}$ by linking each node to its k-nearest neighbors (K-NNs), which are determined based on the position vectors, **pos**. The EdgeConv process then employs a message passing neural network (MPNN) to generate a new feature vector $\mathbf{x}'$ for each node, as defined by the equation

$$\mathbf{x}'_i = \mathop{\square}_{j:(i,j)\in\mathcal{E}} h_{\Theta}(\mathbf{x}_i, \mathbf{x}_j). \tag{1}$$

Here, $h_{\Theta}$ represents a vector function with trainable parameters $\Theta$, and $\square$ denotes an aggregation operation, such as mean or max. Following the MPNN, a new point cloud is formed, denoted as $\mathcal{G}' = \{\mathbf{pos}' \coloneqq \mathbf{x}', \mathbf{x}' \coloneqq \mathbf{x}'\}$. Further operations, like another round of EdgeConv or a graph aggregation operation, can subsequently be applied to $\mathcal{G}'$.

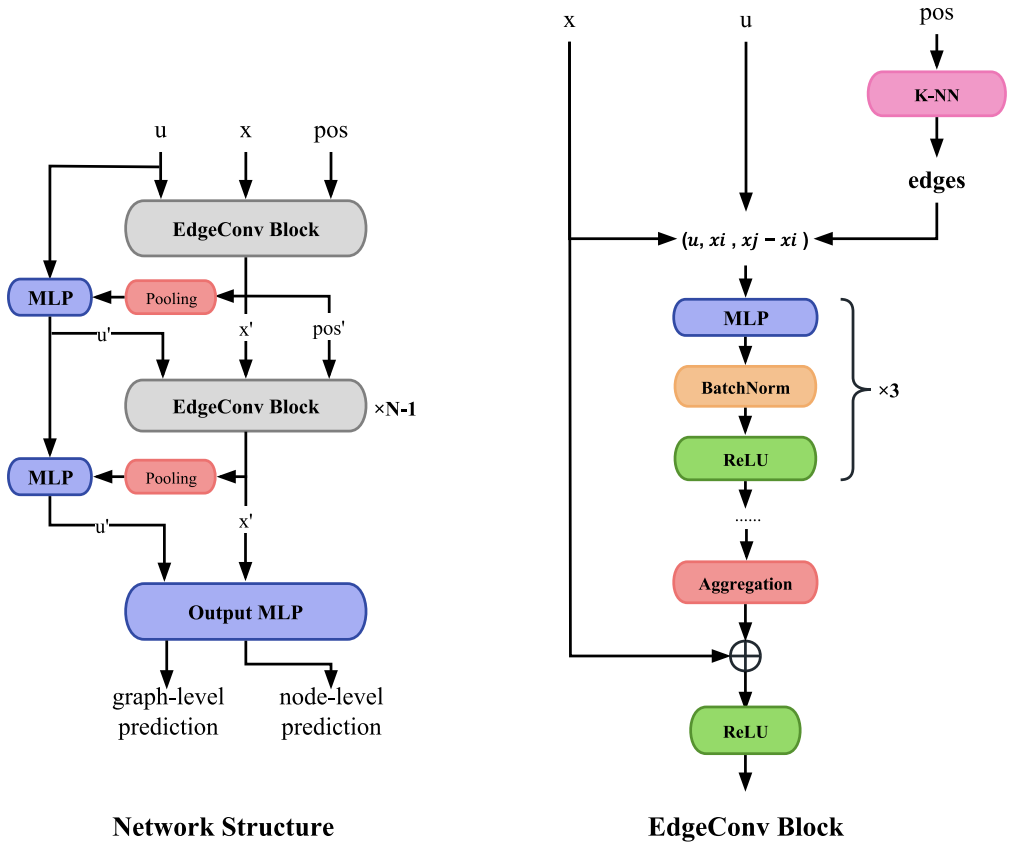

FIG. 2. Network architecture (left) and EdgeConv block (right) used in this study.

EdgeConv operates by convolving information from nearby nodes, effectively functioning as a convolutionlike operation tailored for point clouds. This characteristic makes it particularly suitable for tasks that exhibit typical spatial distribution patterns. In practice, EdgeConv has demonstrated promising results in handling point clouds for various applications, including classification, segmentation [20], and jet tagging [21]. In this study, we adapt EdgeConv to utilize the global information of a point cloud.

The architecture of the GNN used in this study is illustrated in Fig. 2. At the beginning of GNN, node features are normalized using batch normalization [22], followed by the application of a graph aggregation operation which defines a global feature vector $\mathbf{u}$,

$$\mathbf{x} = \text{Batch Normalize}\,(\mathbf{x}), \tag{2}$$

$$\mathbf{u} = g_\Phi\left(\sum_{i\in\mathcal{V}} \mathbf{x}_i/||\mathcal{V}||\right). \tag{3}$$

Here, $||\mathcal{V}||$ represents the number of nodes, and $g_\Phi$ denotes a fully connected (FC) layer with trainable parameters $\Phi$; $\mathbf{u}$ encapsulates the global characteristics by averaging feature vectors across all nodes.

The EdgeConv is then modified from Ref. [20] and ParticleNet [21] to include $\mathbf{u}$ in its operation,

$$\mathbf{x}'_i = \text{ReLU}(\mathbf{x}_i + \underset{j:(i,j)\in\mathcal{E}}{\square} h_\Theta(\mathbf{u}, \mathbf{x}_i, \mathbf{x}_j)). \tag{4}$$

For this, we select $h_\Theta$ to be three successive layers of multilayer perceptron (MLP). Each layer consists of a linear transformation, a batch normalization, and a rectified linear unit (ReLU) [23]. The aggregation function $\square$ is chosen to be max aggregation. A skip connection inspired by Ref. [24] is employed by adding original $x_i$ to the aggregated result. The last ReLU layer introduces nonlinearity and outputs the updated node features $\mathbf{x}'$. For simplicity, the output dimensions of three linear transformations are set to be the same number, then a single scalar can determine an EdgeConv block.

Subsequently, the global feature $\mathbf{u}$ is updated based on the newly obtained node features,

$$\mathbf{u}' = g_\Phi\left(\mathbf{u}, \sum_{i\in\mathcal{V}} \mathbf{x}'_i/||\mathcal{V}||\right). \tag{5}$$

Similar to Eq. (3), $g_\Phi$ is an FC layer with trainable parameters $\Phi$. We denote the combination of the EdgeConv process and the global update process as the update block.

A GNN model can contain several different update blocks. A higher number of blocks enables deeper insight into the graph, at the cost of slower inference speed. After update blocks, an output MLP decodes the node features and generates the final output for the network. The output MLP is a three-layer MLP. The first two layers are both composed of a linear transformation and a ReLU. The output dimension is set to be 256 and 32, separately. The last layer is simply a linear transformation, output dimension equal to the dimension of target vectors. The number of neighbors in the K-NN is set to be 16.

### E. Intermediate variable as truth label

Neural-network-based reconstruction of muon directions typically involves estimating the zenith and azimuth angles $(\theta, \phi)$ or the direction vector $(n_x, n_y, n_z)$ of the muon track. In contrast, this study employs a GNN to predict the positions where the Cherenkov photons are emitted by the muon, an intermediate variable that is further processed by the least-squares method to reconstruct the muon track. It will be demonstrated that this reconstruction pipeline achieves higher accuracy than conventional neural-network-based direction estimation methods.

For each muon event characterized by Monte Carlo truth information—specifying the track's direction and location—we can determine the production position of Cherenkov photons $\mathbf{r}_i$ for the $i$th hDOM at $\mathbf{pos}_i$. The underlying assumption is that the angle between the vector $(\mathbf{pos}_i - \mathbf{r}_i)$ and the muon's directional vector $\hat{n}$ matches the Cherenkov angle. Training the GNN to predict the relative position $(\mathbf{r}_i - \mathbf{pos}_i)$ for each hDOM converts this challenge into a node-level regression task.

For a well-trained model, the sum of the model's prediction, $\mathbf{pred}$, and the hDOM positions, $\mathbf{pos}$, should closely align with the actual muon track, and we parametrize this track with $\mathbf{r}_i = \mathbf{v} + t_i\mathbf{n}$, where $t_i$ represents the photon arrival time of the $i$th hDOM. To estimate the muon's direction vector $\mathbf{n}$, we apply the least-squares method, aiming to minimize the expression $\sum_{i\in\mathcal{V}} w_i(\mathbf{pred}_i + \mathbf{pos}_i - \mathbf{v} - t_i\mathbf{n})^2$. We set the weight of each hDOM $w_i$ to $w_i = \log_{10}(1 + \text{nhits})$, prioritizing hDOMs with more detected photons.

We find that incorporating an intermediate property not only strengthens the method's defense against outliers but also provides a method to assess the quality of event reconstruction. Detailed discussions on the usefulness of intermediate variables are included in Sec. IV C.

## IV. RESULTS AND DISCUSSION

### A. Reconstruction accuracy

The reconstruction performance is measured by two benchmarks: accuracy and run-time efficiency, which are largely determined by two hyperparameters, namely, the number of EdgeConv blocks and the output dimension of each EdgeConv block. Two distinct models are developed to address each of these benchmarks. For real-time event processing, particularly when triggering the signal signature for a neutrino telescope, a "light" model with low computational cost is essential. The light model incorporates three EdgeConv blocks, with output dimension set to 64, 128, and 256, respectively. The total number of trainable parameters in this configuration is 543,877, allowing rapid execution due to the reduced computational complexity. For off-line analysis, where precision is required, the large model is employed, featuring eight EdgeConv blocks. The output dimensions of the first three blocks are 64, 128, and 256, where the remaining five blocks are all 512. Because of the increased complexity, the precision model requires larger computation time.

Both models are implemented with PyTorch Geometric [25,26] and trained on a single 80 GB NVIDIA A100 GPU card. The Adam optimizer [27] is utilized to minimize the weighted mean-squared error loss: $\mathcal{L} = \sum_{\mathcal{G}}[||\mathcal{V}|| \sum_{i\in\mathcal{V}} w_i[\mathbf{pred}_i - (\mathbf{pos}_i - \mathbf{r}_i)]^2 / \sum_{i\in\mathcal{V}} w_i]$, where $w_i = \log_{10}(1 + \text{nhits})$ and $||\mathcal{V}||$ indicates the number of nodes. Weights employed here correspond to the weights used in the least-squares method. The sum of weights of each event is normalized to $||\mathcal{V}||$, in order to reflect the importance in reconstruction for events with more triggered hDOMs. The initial learning rate is set as 0.001 and is reduced by a factor of 2 when the loss plateaus. During training, model parameters are saved at the end of each epoch. The parameters that perform best on the validation dataset are selected for the final model and are then applied to the test dataset, where the results are described next.

The accuracy in predicting the injection positions of Cherenkov photons from the muon track is illustrated in Fig. 3. The distance between reconstructed and truth positions are calculated for each node and its distribution is shown in the plot. As a result, 90% of the prediction error for the large model is smaller than 3 m, and for the light model, it is smaller than 5 m.

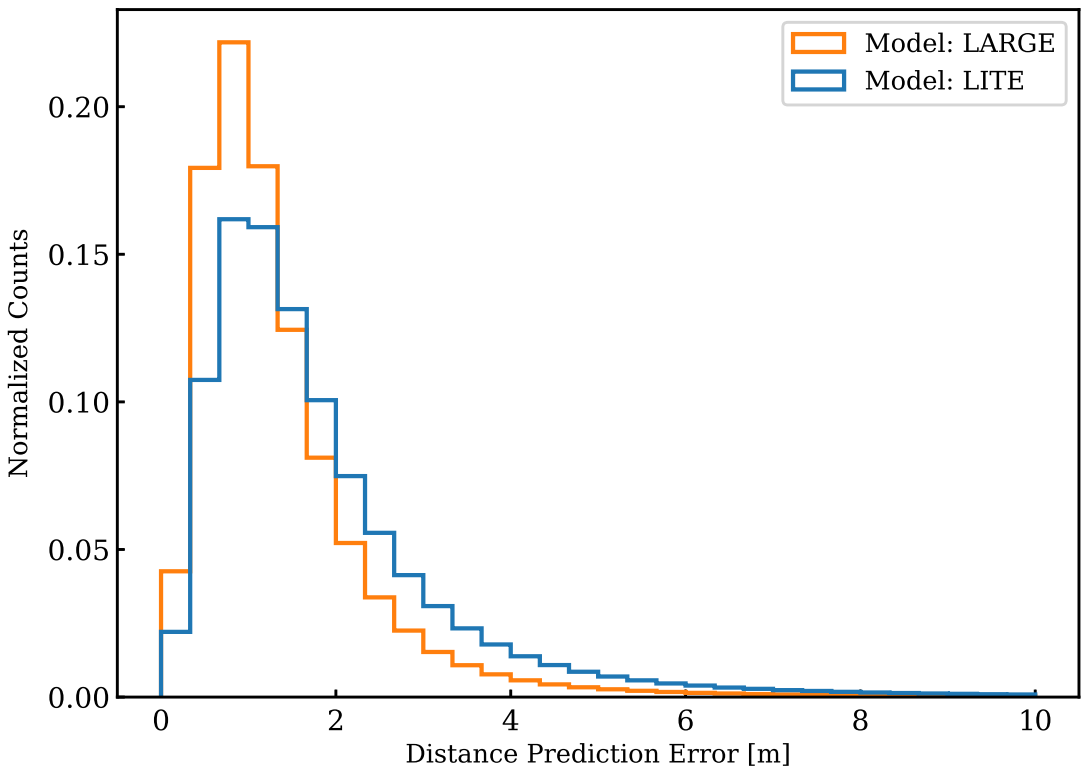

FIG. 3. Distribution of the distance between predicted and true injection positions for the light model (blue) and the large model (orange).

The least-squares method described in Sec. III E is then employed to infer the muon track. Denoting the direction vector of the predicted line as $\mathbf{n}_{\text{pred}}$, the performance of the whole reconstruction chain can be evaluated by angular error, i.e., the angle between $\mathbf{n}_{\text{pred}}$ and $\mathbf{n}_{\text{truth}}$. The median angular error for each energy interval is calculated and the result is illustrated in Fig. 4, where performance of both light and large GNN models are displayed along with the result of likelihood method. For the likelihood method, we utilized the Powell algorithm [28] from the SciPy library [29] to minimize the negative log-likelihood. We configured the hyperparameters to allow an acceptable relative error of $10^{-6}$ for both solution and function convergence, and the maximum allowed number of iterations and function evaluations is set as 10,000. For muons with energies ranging from 100 GeV to 1 TeV, approximately 17.9% of triggered events failed the minimization process. The failure rates for events with energies between 1 and 10 TeV, and between 10 and 100 TeV, are 3.3% and 0.1%, respectively.

An obvious correlation between angular error and energy of the muon can be found in all three methods. A muon with higher energy generally leaves a longer track inside the detector array, which makes the track feature more obvious and thus easier to reconstruct. Moreover, events of lower energy suffer from statistical fluctuation of received Cherenkov photon counts. Therefore, the reconstruction is more accurate for muons with higher energy. The TRIDENT geometry assumes a string length shorter than its total width, such that vertically traveling muons yield shorter tracks compared to horizontal ones. Figure 5 exhibits angular error as a function of $\mathbf{n}_z$. As expected, the accuracy for the horizontal track ($\mathbf{n}_z$ nears 0) is much higher than that of the vertical track ($|\mathbf{n}_z|$ nears 1).

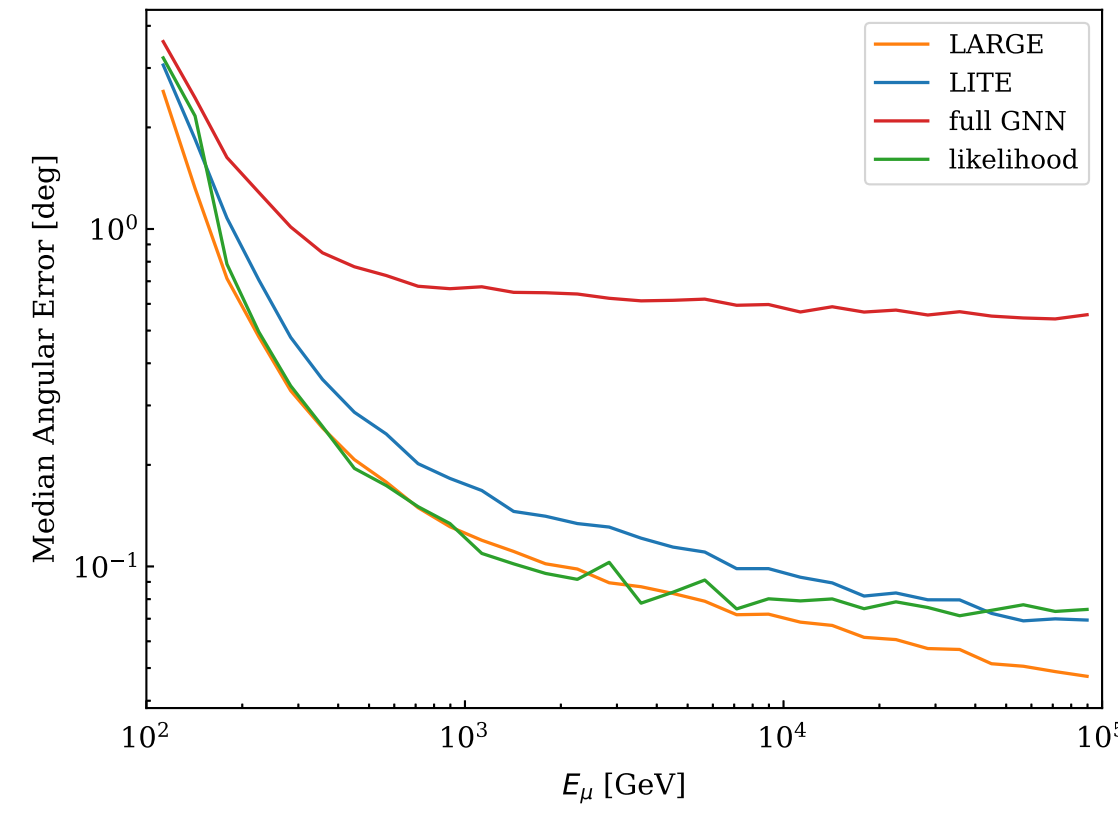

FIG. 4. Median angular error of different methods as a function of muon energy.

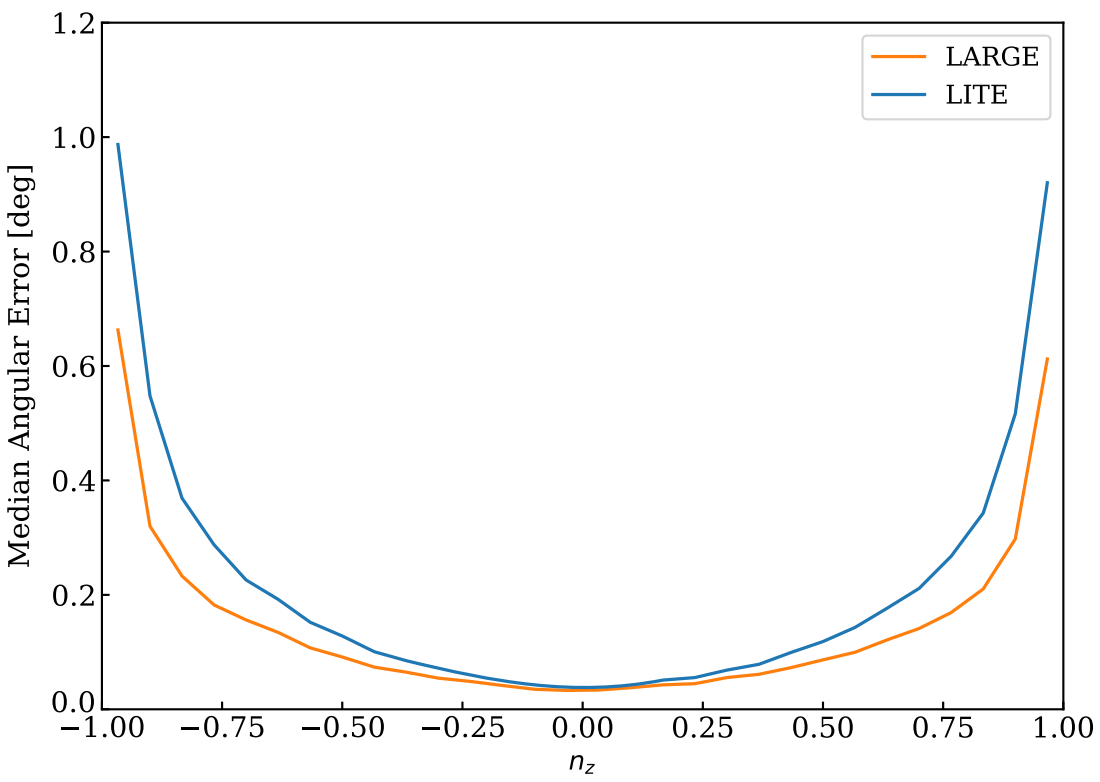

FIG. 5. Median angular error across different models as a function of the cosine of the zenith angle.

### B. Reconstruction time

To evaluate the run-time performance, we measured the inference time per event needed by different methods. Measurement is performed on an Intel Xeon Gold 5220 CPU in a single-thread mode for the CPU test and on an 80 GB NVIDIA A100 GPU for the GPU test. GNN models are implemented with a batch size of 1024 in both methods. The run-time results for all situations are summarized in Table I. For the GNN-based approach, the total run-time reported includes both the GNN and least-squares steps; however, the least-squares computation consists of simple linear operations and contributes negligibly to the overall inference time.

Muon samples are split into three energy regions, 0.1–1, 1–10, and 10–100 TeV, where the three regions are evaluated separately. As mentioned before, a muon event with higher energy is generally easier to reconstruct, and a muon event with higher energy generally contains more recorded photon hits. As a result, it takes longer time for a GNN to perform calculation over all nodes. More photons at high energies make the likelihood easier to converge around the minimum point. Therefore, the high-energy region is slower to reconstruct for a GNN model, but faster for the likelihood method.

With similar reconstruction accuracy, the large model achieves a speed of 3 orders of magnitude faster than that of the likelihood method. At a small cost in accuracy, the light model is even 10 times faster than the large model. With a CPU, the run-time cost of the light model is still 100 times smaller than likelihood method.

Both accuracy and run-time of the GNN model are strongly related to its number of free parameters, such that a suitable model in the real experiment can be trained by considering the trade-off between run-time and accuracy.

### C. Goodness of reconstruction

To effectively utilize the trained model, it is necessary to ascertain the reliability of the reconstructed results before incorporating them into scientific analyses. Certain types of events inherently challenge accurate reconstruction. For instance, events that only trigger hDOMs on a single string cannot have their azimuthal angle ($\phi$) accurately determined. Similarly, corridor events—where muons pass directly through parallel arrays of strings—often leave insufficient detectable photons along the muon track, leading to potentially unreliable predictions. Neural networks, consisting of a series of linear and nonlinear operations, will invariably produce an output, regardless of the complexity or feasibility of event reconstruction. This section explores methods to assess the reliability of reconstructions for the hybrid-GNN approach.

While the loss function used to train the GNN model does not explicitly enforce any constraints, there is an implicit expectation that the points predicted by the model should align roughly along a straight line. Let $\mathcal{D}$ represent the mean distance between the reconstructed track and the points predicted by the GNN, defined as

$$\mathcal{D} \coloneqq \sum_{i\in\mathcal{V}} \mathrm{Distance}(\mathbf{pred}_i, \mathrm{TrackPred})/||\mathcal{V}||.$$

TABLE I. Mean run-time cost per inference.

| Method | Time (0.1–1 TeV) (ms) | Time (1–10 TeV) (ms) | Time (10–100 TeV) (ms) |
|---|---|---|---|
| Likelihood | 1552.30 | 1259.86 | 919.14 |
| GNN light (GPU) | 0.19 | 0.21 | 0.29 |
| GNN large (GPU) | 0.38 | 0.78 | 2.37 |
| GNN light (CPU) | 5.05 | 12.53 | 30.44 |
| GNN large (CPU) | 54.71 | 152.48 | 181.80 |

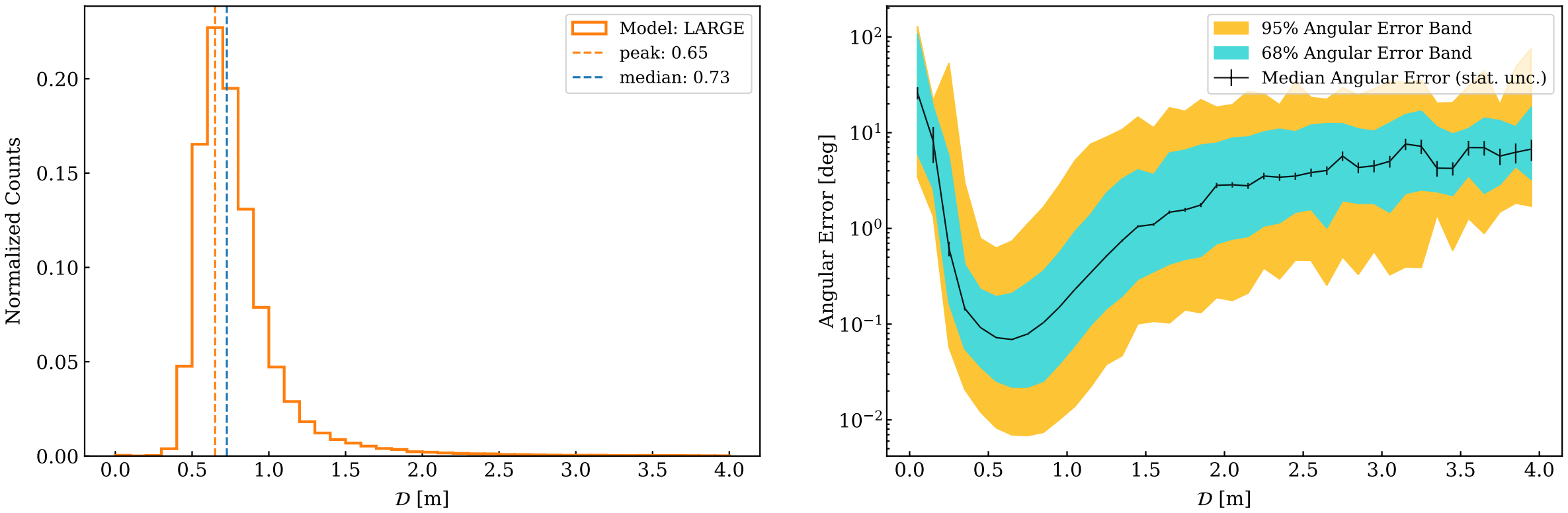

FIG. 6. Left: density distribution of $\mathcal{D}$. Right: angular error as a function of $\mathcal{D}$ with 68% and 95% error bands.

When $||\mathcal{V}||$ is sufficiently large, a small $\mathcal{D}$ suggests a reasonable reconstruction, indicating that the GNN model captures a muon track, predicting a set of points aligned roughly in the track. Conversely, a large $\mathcal{D}$ may signify either poor positional resolution in predicting photon injection positions or a lack of understanding of the events. Thus, $\mathcal{D}$ serves as a natural metric for evaluating the persuasiveness of a prediction.

Illustrating the density distribution of $\mathcal{D}$ for the large model on the left side, Fig. 6 showcases the clustering of $\mathcal{D}$ within a low-value region, indicating the model's good understanding of most muon events. The peak of this distribution can be interpreted as the positional resolution of the GNN model, denoted as $\mathcal{D}_{reso} = 0.65$ m, while the width of the peak reflects the intrinsic uncertainty of the model. On the right side of Fig. 6, the median angular error of the large model, along with the 68% and 95% uncertainty bands, are plotted against $\mathcal{D}$. As expected, the angular error exhibits sensitivity to $\mathcal{D}$, with the minimum angular error observed near $\mathcal{D}_{reso}$. The poorer angular resolution observed in the region with $\mathcal{D} < 0.3$ m predominantly stems from events characterized by small $||\mathcal{V}||$, as will be discussed further below.

The number of triggered hDOMs $||\mathcal{V}||$ significantly influences the efficacy of reconstruction. A small number of triggered hDOMs induces limited available graph information, resulting in unreliable reconstruction. Conversely, a large number of hDOMs often indicates a longer muon track, facilitating clearer track topology and thus easier reconstruction. The density distribution of $||\mathcal{V}||$ and the angular resolution as a function of $||\mathcal{V}||$ for the large model are depicted in Fig. 7. As illustrated in the right plot, the performance of the large model is sensitive to $||\mathcal{V}||$. When the number of triggered hDOMs exceeds 100, 97.6% of reconstruction results (region below the upper edge of orange band) can achieve an accuracy of 0.1°. However, only a small fraction of muon events trigger more than 100 hDOMs, as indicated in the left plot. This diminishes the effectiveness of $||\mathcal{V}||$ compared to $\mathcal{D}$, which exhibits superior resolution in regions with more events and can thus effectively select well-reconstructed areas.

By combining $\mathcal{D}$ and $||\mathcal{V}||$, we gain deeper insights into the quality of reconstruction achieved by the hybrid-GNN method. Illustrated in Fig. 8 is a 2D histogram representing the distribution of $\mathcal{D}$ and the number of triggered hDOMs, alongside their correlation with the 97.6% quantile of angular error. A well-reconstructed region is identified by the 97.6% quantile of angular error values below 5°, highlighted in orange in the right plot. This region contains 87.3% of the reconstructed events.

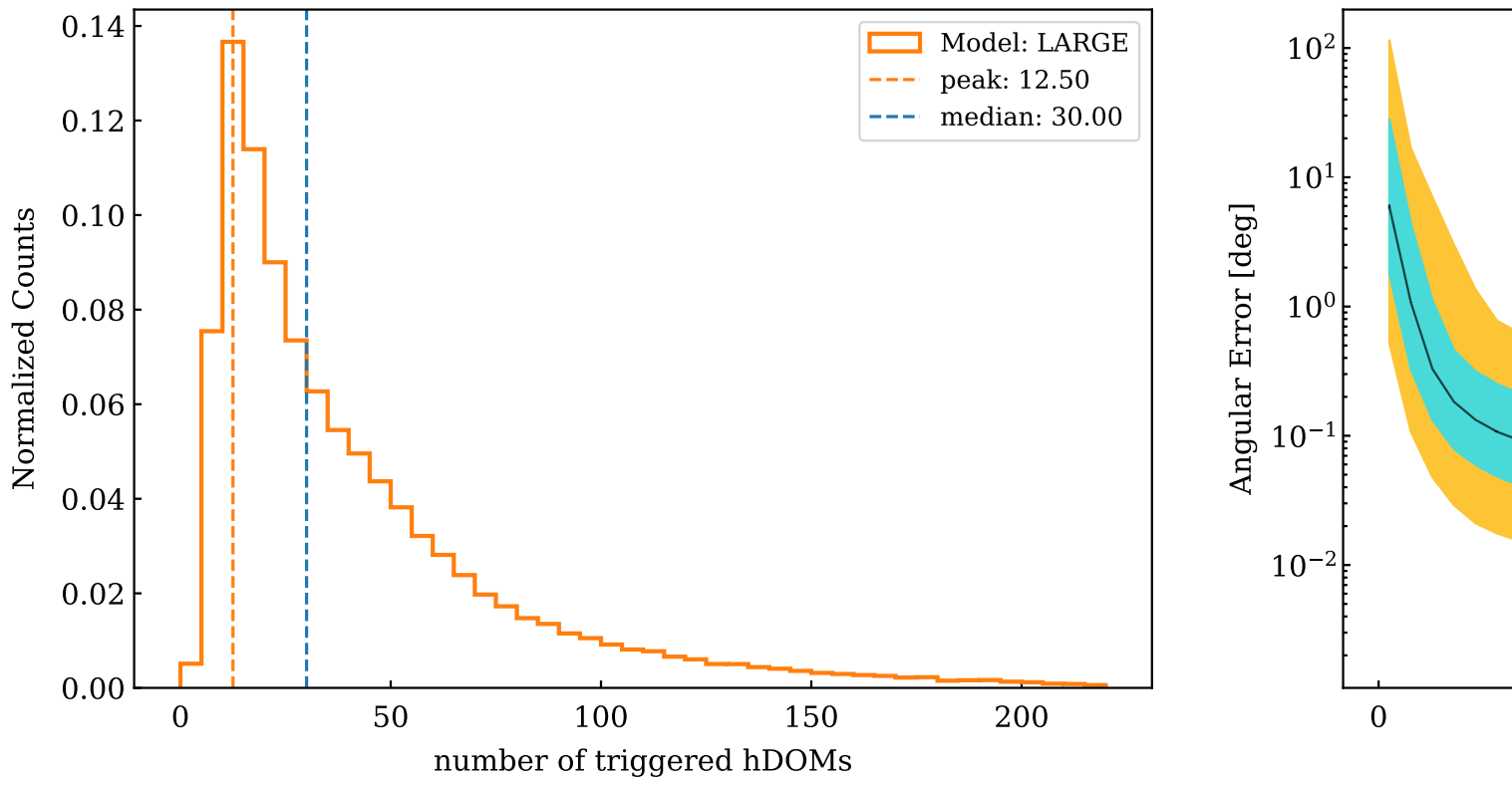

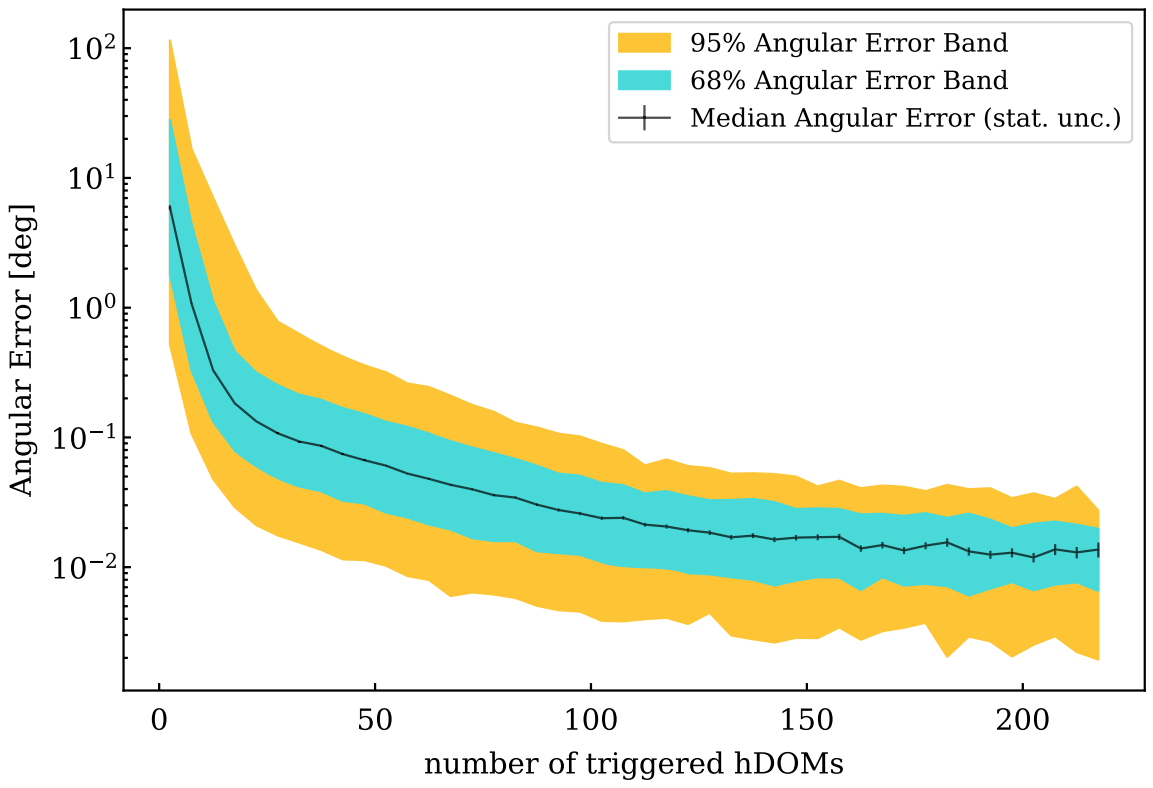

FIG. 7. Left: density distribution of number of triggered hDOMs. Right: angular error as a function of number of triggered hDOMs with 68% and 95% error bands.

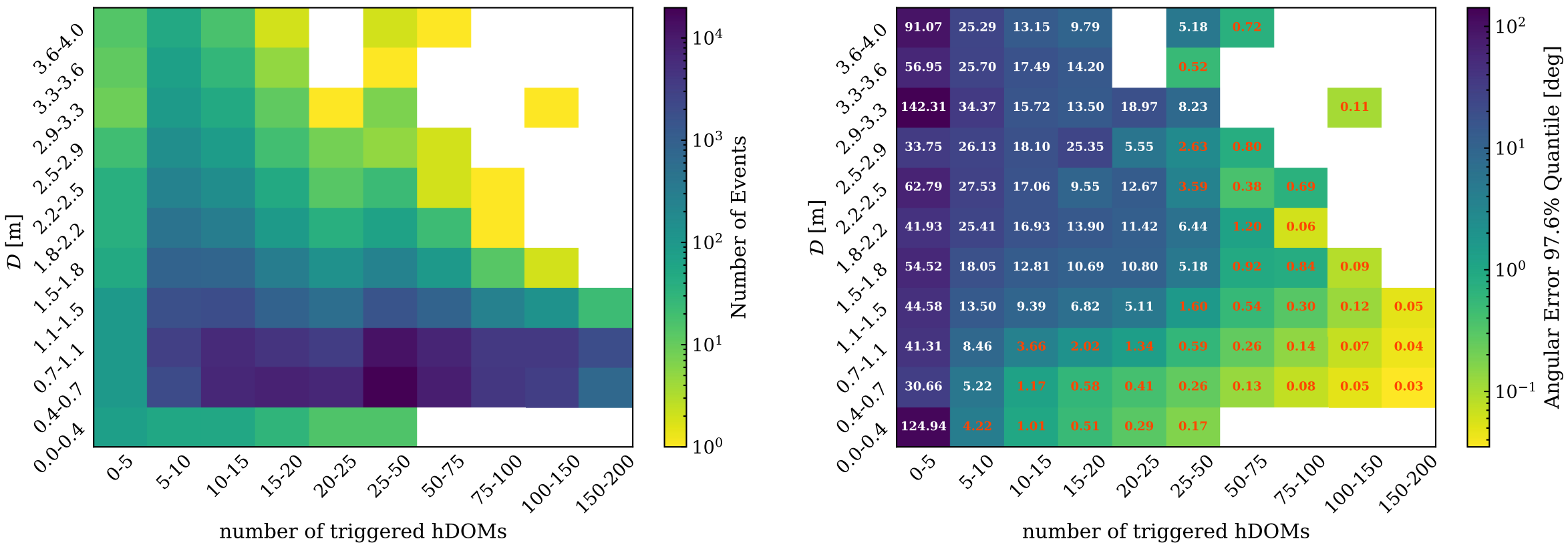

FIG. 8. Left: 2D density distribution of number of triggered hDOMs ($x$ axis) and $\mathcal{D}$ ($y$ axis). Right: angular error 97.6% quantile as a function of number of triggered hDOMs and $\mathcal{D}$. Regions with 97.6% quantile angular error values below 5° are highlighted with orange labels.

This demarcation of regions based on angular resolution enhances sensitivity during actual experiments. Furthermore, the poorly reconstructed region can serve as valuable tool for algorithm calibration.

## V. IMPACT OF $^{40}$K-INDUCED BACKGROUND PHOTONS

Photons produced by the radioactive decay of Potassium-40 ($^{40}$K) constitute the primary source of optical background in deep-sea neutrino telescopes such as TRIDENT. To model the effect of $^{40}$K decays, we use a Geant4-based simulation to construct a template describing the frequency at which each hDOM receives background photons as a function of coincidence level (CL), defined as the number of photons detected within a 20 ns window. The observed rates are approximately 70 kHz for $\mathrm{CL} = 1$, 1 kHz for $\mathrm{CL} = 2$, 0.1 kHz for $\mathrm{CL} = 3$, and less than 0.1 kHz for higher CL. While higher CLs are more signal-like, they are increasingly rare.

This template is then used to simulate the impact of $^{40}$K-induced background photons on muon signal reconstruction. Specifically, for each event, background photons are simulated for the hDOMs located within 200 m of the muon track, with their $t_{\mathrm{res}}$ constrained to the range from −50 to 100 ns. These spatial and temporal criteria are selected because they represent the conditions under which background photons are most likely to mimic signal photons.

GNNs are trained and evaluated on the same simulated events as in the baseline study, but with the addition of background photons from $^{40}$K decay. Figure 9 shows the median angular error as a function of muon energy for each reconstruction method. The presence of background photons degrades the angular resolution for all methods. In the low-energy region, the likelihood-based method is more strongly affected than the GNN-based methods; for example, for 100 GeV muons, the median angular error of the likelihood-based method increases from 3° (without background) to 8° (with background), whereas the GNN-based method shows a smaller increase, from 3° to 4°. At higher energies, the large GNN model performs comparable to the likelihood method. The median angular error remains below 0.1° for muons with energies above 10 TeV.

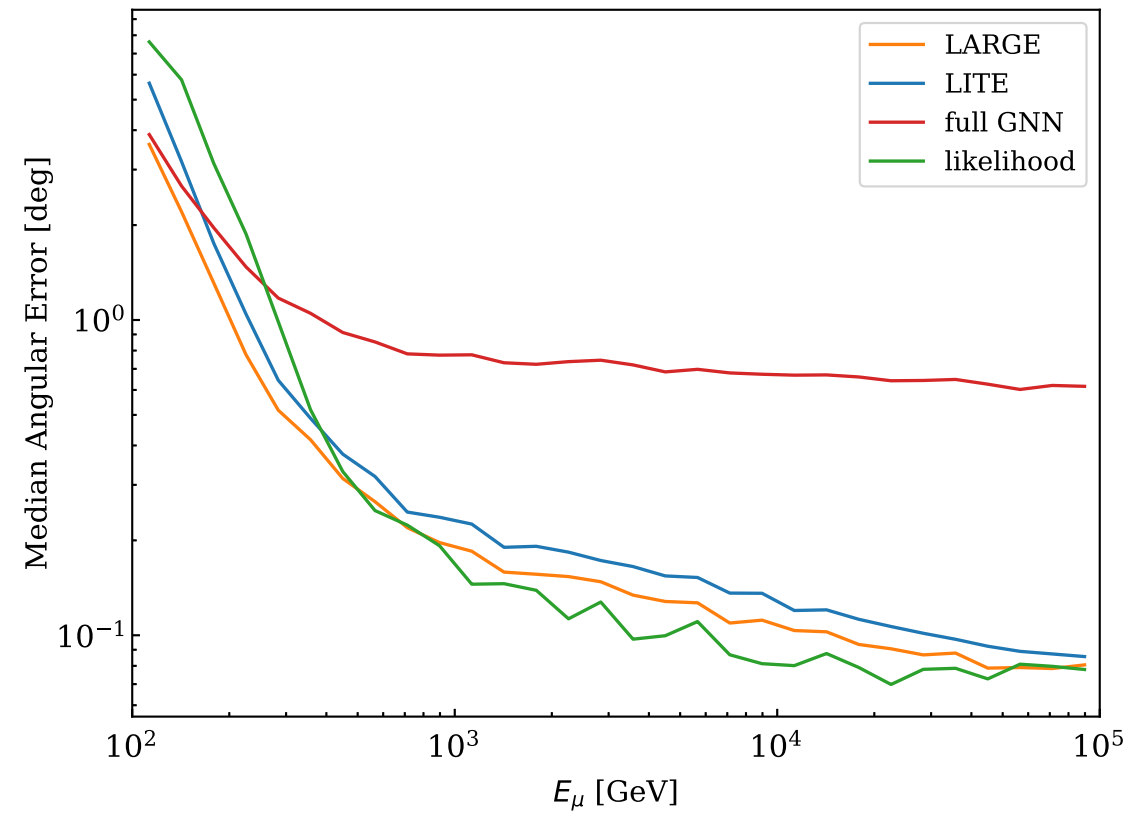

FIG. 9. Comparison of median angular error of different methods evaluated on events with background photons from $^{40}$K decay simulated.

It should be noted that this approach represents a simplified scenario. A comprehensive assessment would require modeling additional background sources, detector electronics, trigger strategies, and noise suppression algorithms. Such studies are planned for future work as the TRIDENT Collaboration refines its experimental strategies.

## VI. SUMMARY

The hybrid-graph neural network method presented in this work offers a significant advancement in the fast and accurate reconstruction of muon tracks in neutrino telescopes. By combining Cherenkov photon injection positions as intermediate variables and leveraging the power of GNN architectures, the method achieves substantial improvements in both speed and accuracy compared to traditional likelihood-based techniques. The light model achieves run times of 0.19–0.29 ms per event on GPUs, which is 3 orders of magnitude faster than the traditional methods assumed here, while maintaining high reconstruction accuracy. The median angular error for high-energy muons (10–100 TeV) is approximately 0.1°. These results demonstrate that the GNN-based method significantly enhances the computational efficiency and precision required for muon reconstruction in neutrino telescopes, making it a robust solution for both real-time event processing (the light model) and off-line high-precision analyses (the large model). Furthermore, the architecture's geometry-independent design ensures scalability and adaptability to various detector configurations, making it highly suitable for next-generation neutrino observatories.

## ACKNOWLEDGMENTS

This work has been reviewed and approved for publication by the TRIDENT Collaboration. All authors are members of the TRIDENT Collaboration and gratefully acknowledge the support, infrastructure, and resources provided by the collaboration. The authors would like to thank Donglian Xu, Fan Hu, Fuyudi Zhang, Hualin Mei, and Weilun Huang for their useful suggestions and discussions during the development of the paper. The authors are also grateful to Jun Guo, Iwan Morton-Blake, and Ziping Ye for their thorough review and valuable feedback for the paper. This work was supported by National Key R&D Program of China (Grant No. 2024YFA1610603) and National Natural Science Foundation of China (Grant No. 12475108).

## DATA AVAILABILITY

The data that support the findings of this article are openly available [30].